\documentclass[prl, showpacs, preprint, floatfix]{revtex4}
\usepackage{latexsym}
\usepackage{amssymb}
\usepackage{amsmath}\large\huge\normalsize
\usepackage{graphicx}
\usepackage{amsfonts}
\usepackage{array}
\usepackage{subfigure}

\newcommand{\be}{\begin{equation}}
\newcommand{\ee}{\end{equation}}
\newcommand{\bea}{\begin{eqnarray}}
\newcommand{\eea}{\end{eqnarray}}

\DeclareMathOperator{\sgn}{sgn}

\begin{document}
\title{Dynamical Singularity Resolution in Spherically Symmetric Black Hole Formation}

\author{Jonathan Ziprick}
\email[Electronic address:]{j.ziprick-ra@Uwinnipeg.ca}

\author{Gabor Kunstatter}
\email[Electronic address:]{g.kunstatter@Uwinnipeg.ca}

\affiliation{Department of Physics and Winnipeg Institute of Theoretical Physics, ${}^*$University of Manitoba and ${}^\dagger$University of Winnipeg, Winnipeg, Manitoba, CANADA}

\received{\today}


\begin{abstract}
We study numerically the effects of loop quantum gravity motivated corrections on  massless scalar field collapse in Painlev\'e-Gullstrand coordinates. Near criticality, the system exhibits Choptuik scaling with the added features of a mass gap and a new scaling relationship dependant upon the quantum length scale. The quantum corrected collapse exhibits a radiation-like phase which resolves the singularity: the black hole consists of a compact region of space-time bounded by a single, smooth trapping horizon. The ``evaporation'' is not complete but leaves behind an outward moving remnant.
\end{abstract}

\pacs{04.25.dc, 04.70.Dy}

\maketitle
\noindent
{\it Introduction.} 
A successful quantum theory of gravity is expected to resolve the singularities present in general relativity. The two leading approaches to quantum gravity are string theory and loop quantum gravity (LQG).  A method developed by Thiemann \cite{Thiemann} to regularize the gravitational potential in LQG has been successful in demonstrating singularity resolution
in mini-superspace models of black holes \cite{Bojowald,Modesto} and quantum cosmologies \cite{Ashtekar}. 
 Moreover, it has long been conjectured that singularity resolution in black hole spacetimes is intimately related to Hawking radiation. Hayward \cite{Hayward} in particular has presented an elegant model of a complete non-singular black hole spacetime in which Hawking radiation at the apparent horizon prevents the formation of a static event horizon. The black hole is replaced by a compact region of trapped surfaces bounded by a single smooth trapping horizon. To the best of our knowledge, neither singularity resolution in LQG nor Hayward's radiative scenario has been explicitly realized in the context of a dynamical field theory.

The purpose of the present letter is to provide such a dynamical realization by examining numerically the effect of a LQG motivated quantum correction to the gravitational potential in the context of spherically symmetric scalar field collapse. We work in Painlev\'e-Gullstrand coordinates because they are regular across apparent horizons and allow for the evolution of trapped surfaces.
The proposed quantum corrections are motivated by the Thiemann trick \cite{Thiemann}: instead of diverging at the origin, the gravitational potential is modified to vanish smoothly within some quantum length scale. The resulting numerical calculations produce a black hole spacetime that is non-singular and has a compact region of trapped surfaces instead of the usual event horizon. In addition, the quantum corrections lead to interesting new properties of Choptuik scaling \cite{Choptuik}, including a mass gap which was also observed recently by Husain \cite{Husain08}.\\[2pt]

\noindent
{\it Equations of Motion.} We begin with a brief statement of the classical equations developed in \cite{ZK}. The dynamical system that we study is that of a collapsing spherically symmetric, massless scalar field in four spacetime dimensions.

The PG metric is given by
\begin{equation}
ds^2 = -\sigma^2dt^2 + \left(dr + \sqrt{\frac{2G{\cal M}}{r}}\sigma dt\right)^2 + r^2 d\Omega^2 ,
\label{reduced metric}
\end{equation}
where $r^2 d\Omega^2$ is the metric on a two sphere of radius $r$. The lapse function $\sigma$ and the Misner-Sharp mass ${\cal M}$ are determined by: 
\begin{eqnarray}
\label{M eq}
{\cal M}^\prime &=&  \frac{1}{2} \left(r^2(\psi^\prime)^2 +\frac{\Pi_\psi^2}{r^2}\right)+\sqrt{\frac{2G{\cal M}}{r}}\psi^\prime \Pi_\psi ,\\
\label{sigma eq}
\sigma^\prime &=& -\sigma \frac{\psi^\prime\Pi_\psi }{2{\cal M}} \sqrt{\frac{2G{\cal M}}{r}},
\end{eqnarray}
where prime denotes differentiation w.r.t. $r$. The dynamical equations for the scalar field $\psi$ and its canonical conjugate $\Pi_\psi$ are:
\begin{eqnarray}
\label{psi dot}
\dot{\psi} &=& \sigma \left(\frac{\Pi_\psi}{r^2}+\sqrt{\frac{2G{\cal M}}{r}} \psi^\prime \right),\\
\label{Pi dot}
\dot{\Pi}_\psi &=& \left[\sigma \left( r^2\psi^\prime + \sqrt{\frac{2G{\cal M}}{r}} \Pi_\psi \right) \right]^\prime.
\end{eqnarray}

These equations need to be supplemented by boundary conditions for ${\cal M}$ and $\sigma$. Without loss of generality we scale the time coordinate by choosing $\sigma(t,0)=1$.
We also fix ${\cal M}(t,0)=0$ which guarantees a flat metric in the neighbourhood of the origin. A non-zero value of ${\cal M}$ at the origin signals the formation of a singularity.

We now motivate quantum corrections that may arise due to the underlying spatial discreteness implied by LQG. We look to modify terms deriving from the potential which may diverge at the origin. These are easily identified as the  ${2G{\cal M}}/{r}$ terms in Eqs.(\ref{M eq}) to (\ref{Pi dot}). The ${\Pi_\psi}/{r^2}$ terms are kinematical and do not diverge.

Polymer quantization \cite{Polymerization} as used in LQG mini-superspace models replaces the momentum $p$ conjugate to $x$ by the finite translation operator  $U_\mu=e^{i\mu p}$. The quantum version $\widehat{U}_\mu=e^{i\mu \hat{p}}$ has an action given by:
\begin{equation}
(\widehat{U}_\mu \psi)(x) = \psi(x+\mu). 
\label{Uact}
\end{equation}
For the moment we  consider $x$ to span the entire real line. The Thiemann trick \cite{Thiemann} replaces the classical divergent function $f_{unreg}=\sgn(x)/\sqrt{|x|}$  by a commutator:
\begin{eqnarray}
\nonumber
f_{unreg}(x) &\rightarrow& \frac{1}{i\mu}\widehat{U}_\mu^\dagger \left[ \sqrt{|x|},\widehat{U}_\mu \right] - \frac{1}{i\mu}\widehat{U}_\mu \left[\sqrt{|x|},\widehat{U}_\mu^\dagger, \right];\nonumber\\
\label{FD:reg}
                          &\rightarrow&f_{reg}(x)= \frac{\sqrt{|x+\mu|}-\sqrt{|x-\mu|}}{\mu}.
\end{eqnarray}
Notice that $f_{reg}$
is well defined at the origin but has a cusp at $x=\mu$. As in \cite{Husain08}, we instead use a smoothed function that has the same qualitative features as $f_{reg}$: it is nearly equal to $f_{unreg}$ for $x\gg \mu$, it has a positive slope for $x<\mu$ and it is equal to zero at $x=0$.
\begin{equation}
\label{fsmooth}
f_{smooth}(x) =  \frac{\sgn(x)}{\sqrt{|x|}}\sqrt{1-e^{-\left(\frac{x}{\mu}\right)^2}}.
\end{equation}

\begin{figure}[htb]
\begin{center}
\includegraphics[width=3in]{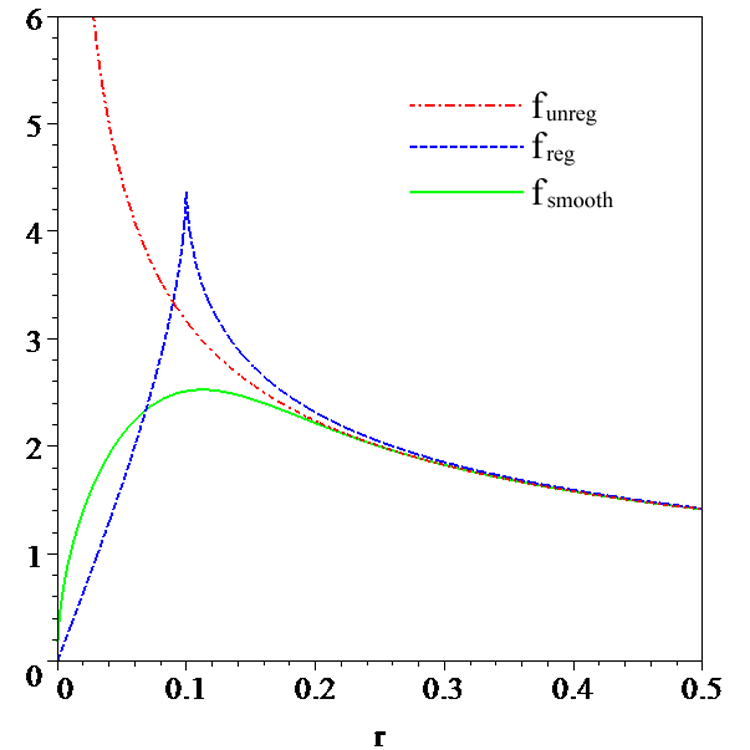}
\caption[$f_{unreg}(r)$, $f_{reg}(r)$ and $f_{smooth}$(r)]{(Color online.) A comparison of $f_{unreg}(r)$, $f_{reg}(r)$ and $f_{smooth}$(r) with $\mu=0.1$.}
\label{jreg}
\end{center}
\end{figure}

Since the equations of motion depend on $r\ge0$ (rather than $x$) we revert to the radial coordinate. See Fig.~\ref{jreg} for a comparison of $f_{unreg}$, $f_{reg}$ and $f_{smooth}$ in this domain.
The quantum correction
is simply the replacement
\begin{equation}
\sqrt{\frac{2G{\cal M}}{r}} \rightarrow \sqrt{2G\cal{M}} f_{smooth}(r).
\end{equation}
in each of (\ref{M eq}--\ref{Pi dot}). These corrections effectively add a repulsive component to the gravitational potential very near the origin, which is  consistent with the expectation that the underlying discreteness implied by LQG at the microscopic level will give rise to a short range repulsion in the semiclassical limit.\\[2pt]

\noindent
{\it Numerical Methods.}
Our code is essentially identical to that used in the recent investigation of Choptuik scaling in PG coordinates \cite{ZK}, where further details can be found. One must first specify on a discrete spatial grid the initial data for the scalar field. Two forms were considered:
\begin{eqnarray}
\label{tanh}
\psi &=& A r^2 \exp{\left[-\left(\frac{r-r_0}{B}\right)^2\right]},\\
\label{gaussian}
\psi &=& A \tanh{\left(\frac{r-r_0}{B}\right)},
\end{eqnarray}
with $\Pi_\psi(r,0)=0$ in all cases.
$A$, $B$ and $r_0$ are parameters which may be varied to study 
critical collapse.

Equations (\ref{M eq}) and (\ref{sigma eq}) determine the mass and lapse functions which are inserted into the right hand side of the time evolution equations (\ref{psi dot}) and (\ref{Pi dot}) in order to calculate the scalar field and its conjugate at the next time slice. These are then used as initial data for the subsequent evolution. This process is repeated until either a singularity forms or the matter disperses to leave behind a flat spacetime.  

In order to adequately resolve the small $r$ structure 
we refined the $r$-spacing $\Delta r(r)$ near the origin. The code destabilized if information moved over too many $r$-points in a single time step, so we used an adaptive time step $\Delta t(t)$ refinement 
\begin{equation}
\Delta t(t) = \hbox{MIN}_r\left\{\frac{dt}{dr} \Delta r(r)\right\} ,
\end{equation}
where $\frac{dt}{dr}$ is the inverse of the local speed of an ingoing null geodesic and $\hbox{MIN}_r$ denotes the minimum across the time slice. \\[2pt]

\noindent
{\it Quantum Corrected Choptuik Scaling.}
Initial data can be separated into two regions separated by a critical value, $p_*$, of some parameter $p$. If the ADM mass increases with increasing $p$, a black hole forms for $p>p_*$ with the inevitable singularity classically. If $p<p_*$, no horizon forms and all matter disperses to infinity.
If $p\sim p_*$, the solution temporarily echoes the critical solution that describes a discretely self-similar zero mass black hole. The existence of this critical solution as an intermediate attractor gives rise to the universal mass scaling properties first observed by Choptuik \cite{Choptuik}. 

\begin{figure}[htb!]
\begin{center}
\includegraphics[width=4in]{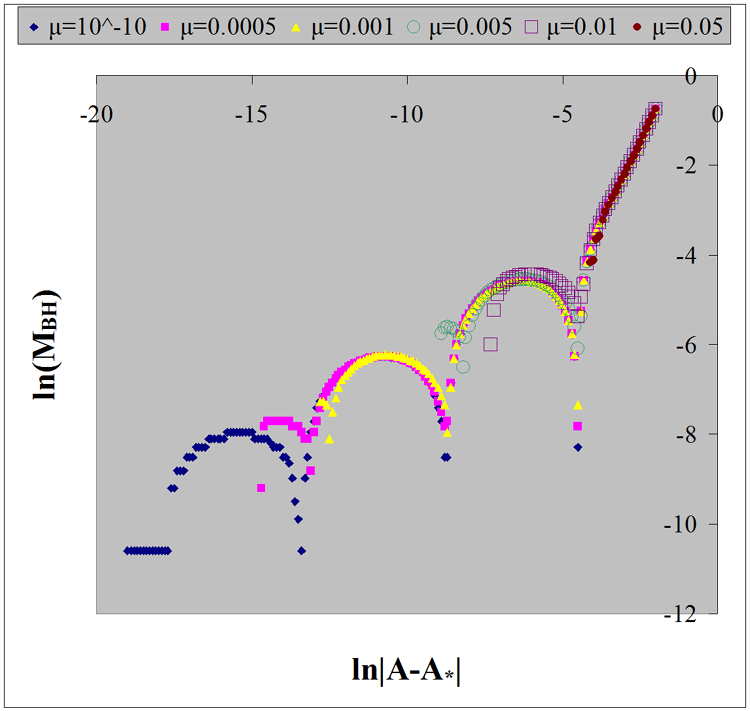}
\caption{(Color online.) Plots of the mass scaling behaviour for a variation in the parameter $A$ in gaussian initial data on a mesh with $\Delta r=10^{-4}$ near the origin.}
\label{qmass}
\end{center}
\end{figure}

Our quantum mass scaling results
are given in Fig.~\ref{qmass}.
These curves in PG coordinates are non-sinusoidal with large amplitude cusps \cite{ZK}. The period and slope are in excellent agreement with previous studies of Choptuik scaling despite the unusual functional form. The new feature in the present case is a change in the critical parameter (which we refer to as $p_{q*}$) that is dependant upon the choice of quantum length scale $\mu$. For sufficiently small $\mu \ll \Delta r(0)$, the system is numerically identical to the classical case (see the $\mu=10^{-10}$ line in Fig. \ref{qmass}).  For larger $\mu$ (though still small enough to be near criticality) the solutions exhibit critical scaling for some range $p>p_{q*}$ and come to an abrupt stop at $p=p_{q*}$, below which no black holes form. This is the mass gap predicted by quantum gravity \cite{Bojowald}, and recently found by Husain \cite{Husain08} using a numerical model in null coordinates with similar corrections to those used here. It is interesting that in PG coordinates larger $p_{q*}$ does not necessarily correspond to larger black hole mass due to the non-monotonic, large amplitude oscillations in the scaling relation.

\begin{figure}[htb]
\includegraphics[width=3in]{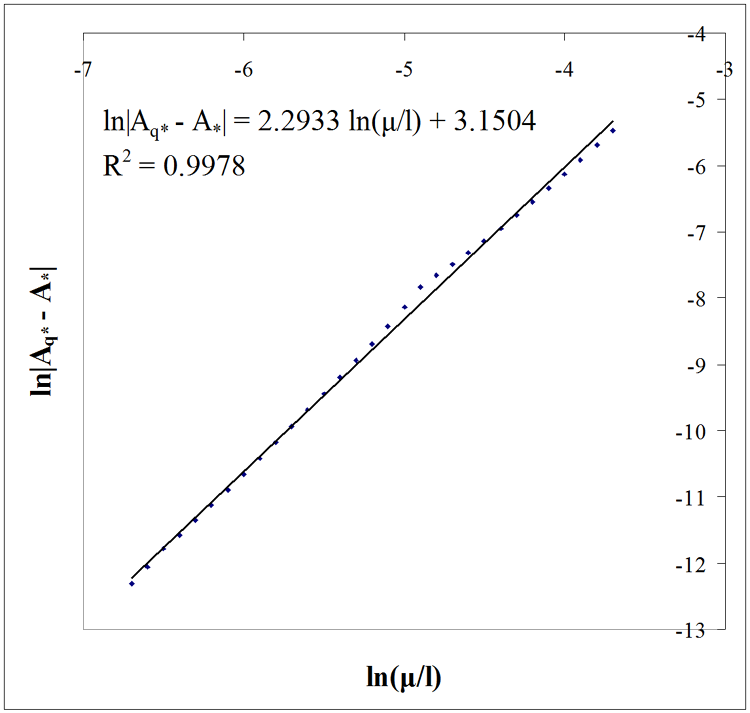}
\caption{\label{AvsL}
Variation in the parameter $A$ in tanh initial data.
}
\end{figure}

Remarkably, we find near criticality a quantum power scaling relation:
\begin{equation}
p_{q*}=k_p\left(\frac{\mu}{l}\right)^\beta + p_*,
\end{equation}
where
$k_p$ is a family dependant constant, $l$ is a length parameter required for proper units and $\beta=2.29\pm0.04$ is a universal constant whose value was verified by varying $A$ and $B/l$ in both forms of initial data.
See Fig.~\ref{AvsL} for an example.\\

\begin{figure}[htb]
\centering
\subfigure[Classical]{
\includegraphics[width=0.4\linewidth]{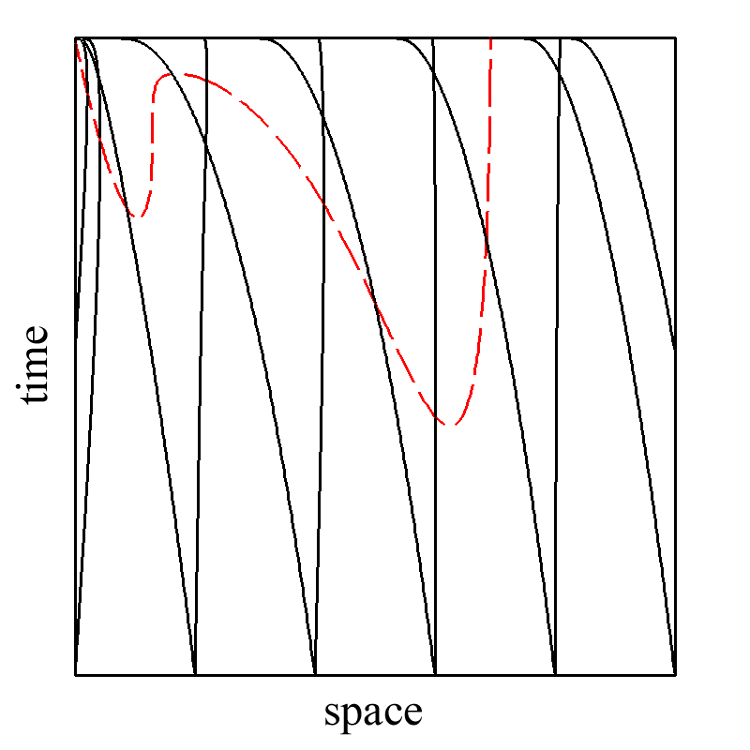}
\label{qm1}
}
\hspace{0.25in}
\subfigure[Quantum]{
\includegraphics[width=0.4\linewidth]{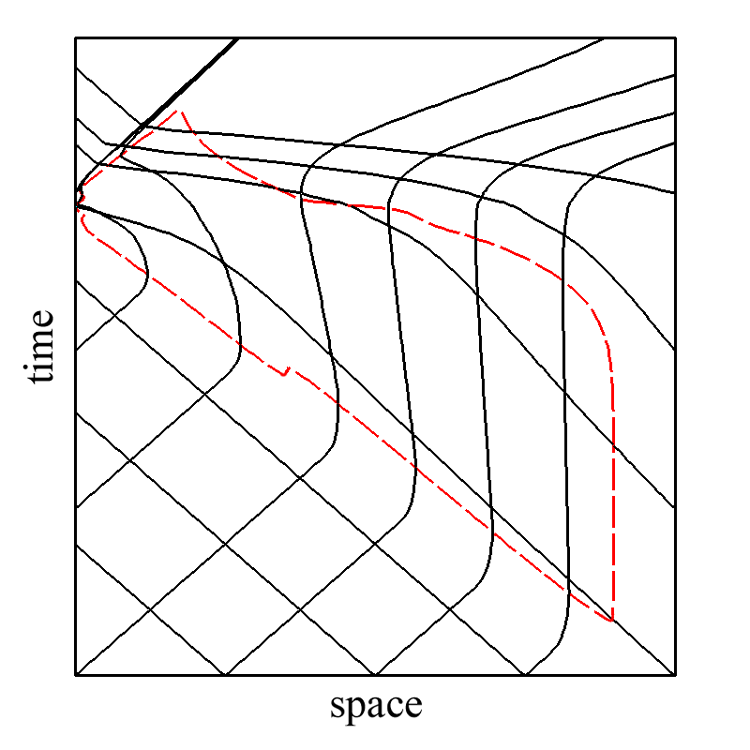}
\label{qm2}
}
\caption{\label{qm}(Color online.) Spacetime diagrams. The (black) solid lines are null geodesics and (red) dashed lines indicate the trapping surface boundary.}
\end{figure}

\noindent
{\it Singularity Avoidance.} Figure \ref{qm} provides spacetime diagrams for gaussian initial data far from criticality ($p\gg p{_q*}$) for the classical and quantum corrected potentials. Animations of the time evolution of the mass density corresponding to Figs. \ref{qm1} and \ref{qm2} are available for viewing at \url{http://theoryx5.uwinnipeg.ca/users/jziprick/} . 
In both cases inner and outer apparent horizons
(where outward moving radial null rays are stationary) appear at some initial PG time and then separate as the inner horizon moves inward. In between the two horizons exist trapped surfaces on which all null rays move inward. For generic initial data the outer horizon moves outward until the last of the matter has fallen through it.
Interestingly, a second pair of horizons forms for a period in both cases before annihilating to leave a single pair.

In the classical case, the singularity forms when the inner horizon reaches the origin, at which point the code terminates. The outer horizon remains stationary once it has reached its maximum radius; it cannot decrease because the scalar field obeys the standard energy conditions.

Figure \ref{qm2} clearly exhibits singularity avoidance in the quantum case.  As expected, the outer horizon $r_{oh}$ reaches a maximum radius after all mass in the vicinity has passed through. As the inner horizon nears the quantum length scale it begins to slow down, eventually bouncing and moving outward without ever reaching the origin. Near the bounce, the
mass density takes on negative values. This leads to mass loss and causes the outer horizon to shrink until it meets the inner horizon in annihilation, leaving behind an outward moving remnant with strictly positive mass density. The amount of mass lost during the non-conservative ``evaporation'' phase seems numerically to be arbitrarily small for horizons forming in the quantum region which evaporate very soon after formation (for $\mu/r_{oh,max} \sim 1$). For larger black holes ($\mu/r_{oh,max} \sim 1/5$) we observed mass losses of up to 80\%, and higher values are likely possible. One can speculate that the quantum corrected collapse of macroscopic black holes would also leave microscopic remnants with radii of order $\mu$.

According to Einstein field equations, the energy density measured by an observer at constant $r$ is
\begin{equation}
\rho = \frac{{\cal M}^\prime}{4 \pi r^2},
\end{equation}
The negative mass density near the quantum bounce thus implies a violation of the null and dominant energy conditions inside the black hole. Such violations are also expected \cite{Visser} in the presence of Hawking radiation.

For a time after the outer horizon reaches its maximum and before it starts moving inward, the spacetime appears from the exterior to be that of a static black hole. The amount of time that the outer horizon remains stationary depends upon the distance the mass has to travel from $r=r_{oh}$ down to the quantum region near $r \sim \mu$. The shape of the mass distribution is also a factor since even for $r_{oh} \gg \mu$, the outer horizon may continue to grow until the onset of evaporation if the infalling mass is spread out over a distance of the same magnitude or greater than $r_{oh}$. By adjusting the initial mass distribution and $\mu$, the lifetime of the seemingly classical black hole can be made arbitrarily large or small. 

In the study by Hayward \cite{Hayward}, a small length parameter was used to effectively preclude singularity formation. A Vaidya-like  region with positive ingoing energy flux was matched to a stationary black hole spacetime to model black hole formation. After an arbitrary ``time'' given by the null coordinate $v$, a Vaidya-like region with negative ingoing energy flux was matched to the stationary black hole to model evaporation. The negative energy flux was balanced by outgoing Hawking radiation in order to conserve energy. This scenario bears a striking similarity to the black hole evolution produced by our numerical simulations, suggesting that the disappearing mass in our system is in some sense being radiated away, despite the fact that we have not included Hawking radiation directly. The key point in our analysis is that the quantum correction contains a repulsive core that prevents the inner horizon from reaching the origin, so that the singularity cannot form. The subtle (non-local) interplay between the microscopic behaviour of the potential and the dynamics of the outer horizon, accompanied by violations of the energy conditions, results in acausal behavior of the semiclassical system.  Moreover, since there is a mass remnant left behind after ``evaporation'', our semiclassical scenario has more than one feature relevant to the ultimate resolution of the information loss paradox. Although many questions remain, our analysis provides tantalizing support for singularity avoidance in dynamical black hole formation.\\[2pt]


\begin{acknowledgments}
We thank R. Daghigh, D. Garfinkle, C. Gundlach, J.M. Martin-Garcia, V. Husain, R. Kobes, K. Lake, J. Louko and H. Maeda and A. Peltola for helpful discussions and correspondence. The authors are grateful to Westgrid for providing the computer resources and the Natural Sciences and Engineering Research Council of Canada for financial support.
\end{acknowledgments}

\end{document}